# Low-frequency Noise in Individual Carbon Nanotube Field-Effect Transistors with Top, Side and Back Gate Configurations: Effect of Gamma Irradiation

Sydoruk V A[1], Goß K[1,a], Meyer C[1], Petrychuk M V[2], Danilchenko B A[3], Weber P[1,4], Stampfer C[1,4], Li J[1] and Vitusevich S A[1,*]

[1]Peter Grünberg Institute (PGI-6/8/9; IBG-2) and Jülich-Aachen Research Alliance for Future Information Technology (JARA-FIT), Forschungszentrum Jülich, 52425 Jülich, Germany
[2]Taras Shevchenko National University, 03022 Kyiv, Ukraine
[3]Institute of Physics, NASU, 03028 Kyiv, Ukraine
[4]JARA-FIT and II. Institute of Physics, RWTH Aachen University, 52074 Aachen, Germany
[a]Current address: Physikalisches Institut, Universität Stuttgart, Germany

**Abstract** We report on the influence of low gamma irradiation ($10^4$ Gy) on the noise properties of individual carbon nanotube (CNT) field-effect transistors (FETs) with different gate configurations and two different dielectric layers, $SiO_2$ and $Al_2O_3$. Before treatment, strong generation-recombination (GR) noise components are observed. These data are used to identify several charge traps related to dielectric layers of the FETs by determining their activation energy. Investigation of samples with a single $SiO_2$ dielectric layer as well as with two dielectric layers allows us to separate traps for each of the two dielectric layers. We reveal that each charge trap level observed in the side gate operation splits into two levels in top gate operation due to a different potential profile along the CNT channel. After gamma irradiation, only reduced flicker noise is registered in the noise spectra, which indicates a decrease of the number of charge traps. The mobility, which is estimated to be larger than $2 \times 10^4$ $cm^2V^{-1}s^{-1}$ at room temperature, decreases only slightly after radiation treatment demonstrating high radiation hardness of the CNTs. Finally, we study the influence of Schottky barriers at the metal-nanotube interface on the transport properties of FETs analyzing the behavior of the flicker noise component.

**PACS: 05.40.Ca; 85.30.Tv; 77.55.-g; 81.05.U-; 61.48.De.**

[*] Corresponding author. E-mail: s.vitusevich@fz-juelich.de. On leave from Institute of Semiconductor Physics, NASU, Kyiv, Ukraine.

## I. INTRODUCTION

Carbon nanotubes (CNTs) are a promising material for future nanoelectronics. One of their applications is in field-effect transistors (FETs). Since the first demonstration in 1998 by Tans et al.[1] , CNT-FETs have attracted increasing attention and significant progress has been made in understanding their operation and improving device performance.[2–4] However, developing an ideal CNT-FET includes the variation of parameters such as the contact metallization, the chirality of the CNT, the gate dielectric material and the geometry of the gate electrodes.[4–10] In addition, the design and fabrication of the contacts play an important role. CNT-FETs have generally relied on direct metal-semiconductor contacts, usually forming Schottky barriers.[7,11–13] Heinze et al.[14] showed that CNT-FETs function as "unconventional Schottky barrier transistors", where the the gate voltage is primarily tuning the contact resistance rather than the channel conductance. Nevertheless, it is still unclear how the interface morphology, the area of the contact region and the properties of the contact barriers influence the transport mechanisms.[15–17]

Gamma irradiation is widely used to modify various characteristics of different materials. For example, an improvement of DC and RF electrical performance after a cumulative gamma radiation dose of $4.28 \times 10^2$ Gy and high radiation hardness up to a radiation dose of $8.5 \times 10^3$ Gy has been shown for GaAs-based FETs.[18,19] The improvement of the transistor functionality using gamma irradiation treatment is mainly due to the formation of more homogeneous contacts, a decrease in structural defects and stress relaxation effects.

Concerning individual CNTs, up to now, only high radiation doses greater than $5 \times 10^4$ Gy have been used to tune the characteristics of CNT-based devices.[20,21] No studies on the effect of small radiation doses on the properties of individual CNT-FETs and their radiation hardness have yet been reported. Our previous results obtained on multiple parallel-aligned CNT-FETs indicate an improvement of the transport characteristics, such as a decreasing leakage current after gamma irradiation of samples with small doses.[22] Additionally, we demonstrated that the high radiation hardness of CNTs allows us to use gamma radiation treatment to modify the contacts and $SiO_2$ dielectric layer without modifying the CNT material properties.[23] We analyzed CNT-FET performance by investigating their noise properties. Different operating regimes of CNT-FETs are compared[22,23], allowing us to identify the ideal operating point for different purposes in future devices. This is especially useful for devices which require complex gate structures, as for instance multiple serial quantum dots in CNTs. Comparing devices where the CNTs are fully covered by dielectric material[24] with devices where only a small fraction of the CNTs are covered with an oxide[25–28] shows that charge fluctuations interfering with the quantum transport are much stronger in the former. This indicates that charge traps in the dielectric material play a significant role in device performance.

The Hooge parameter $\alpha_H$, is usually used to compare the noise level of different materials. Initially, $\alpha_H$ was introduced as a constant (~$2 \times 10^{-3}$) for any material[29], but later its dependence on different scattering mechanisms was shown.[30] In the case of CNTs, the value of the Hooge parameter depends on the performance of the dielectric layers between the gate electrode and the nanotube, and on the CNT's environment. Additionally, a passivation of the CNT can influence the noise properties.[31–33] It should be noted that for CNTs $\alpha_H$ was shown to be in the range from $10^{-5}$ to 1, with the most probable value of the parameter being about $10^{-3}$.[31–36] In addition to flicker noise, random telegraph signal (RTS) noise is generally observed in CNT-FETs. In this case, trapping-detrapping processes occur in the occupancy of individual charge traps resulting in strong fluctuations of the channel current with random discrete impulses of equal height. Along with the two distinct levels usually registered, multilevel impulses may be observed on the current time trace. The appearance of RTS noise is related to the existence of individual defects/traps in the dielectric layer close to the CNT transport channel.[37–40] RTS noise usually dominates all other noise sources, including $1/f$ and generation-recombination (GR) noise components and corresponds to a non-Gaussian distribution in a noise signal. At the same time, GR noise represents a noise source with Gaussian distribution of instantaneous values of current/voltage due to fluctuations in a number of channel carriers associated with random transitions of charge carriers between energy bands and trap states located near the Fermi level of the channel. The GR noise components resolved in the noise spectra contain information about the particular properties of GR processes: magnitudes of free carrier lifetimes and the levels and bands between which these processes occur.[41] These data can be used to analyze the parameters of the main traps that determine the transport properties of the FET channel, as in the case of the deep-level transient spectroscopy (DLTS) technique.[42]

In the present work, we investigate the electronic properties of CNT-FETs with different gate configurations and dielectric layers by applying noise spectroscopy techniques. Using $SiO_2$ and $Al_2O_3$ dielectric layers for the CNT transistors and analyzing the generation-recombination noise components, we identify the trap energies, separately related to the two dielectric layers. These traps are responsible for the fluctuation processes in the transport channel. We investigate differences in the influence of side and top gate on active traps in the transistor structures. Exposure of the FET devices to gamma radiation has an impact on the electronic transport and the noise properties: in spite of a slightly decreased carrier mobility, the total current in the on-state and the maximum value of transconductance increase, while the levels of the flicker and GR noise components decrease at the same time.

## II. EXPERIMENTAL DETAILS AND DEVICE FABRICATION

We fabricated and investigated FETs based on individual carbon nanotubes with top and side gate configurations. The CNTs were grown on Si/SiO$_2$ substrates by chemical vapor deposition (CVD) using Fe/Mo as catalyst and methane as the carbon feed gas.[43] The thickness of the SiO$_2$ layer was 200 nm. Using atomic force microscopy (AFM), the position of the CNTs can be determined relative to pre-patterned marker structures. Before the fabrication of the electrodes, the CNTs were covered with a layer of Al$_2$O$_3$ (25 nm) by pulsed laser deposition (PLD) in the region designated for the gate structures. All electrodes were patterned in one final step forming source and drain contacts in the uncovered CNT region and side and top gate electrodes in the region where the CNT was covered by the dielectric layer. The transport and noise measurements presented here were taken on two of the devices. An AFM image of one of the CNT devices with different gate configurations is shown in Fig. 1. The diameter of the CNT in Fig. 1 is estimated by AFM to be in the range of 1.7-2 nm. Assuming a semiconducting CNT we can estimate its energy gap to be in the range of 0.35-0.40 eV.[44,45] The semiconductor features of the CNTs were checked by electric measurements. Palladium was used for the contact and gate electrodes since it has a relatively large electron work function in the range of 5.22-5.6 eV. This resulted in almost transparent Schottky barrier contacts to the CNTs.[13]

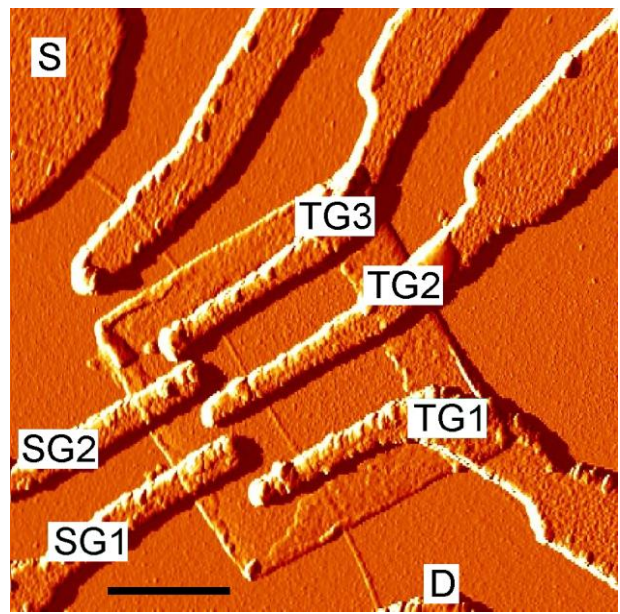

Fig. 1. AFM image of the investigated CNT-FET with source (S) and drain (D) contacts, as well as two side (SG1, SG2) and three top (TG1, TG2, TG3) gates. The Al$_2$O$_3$ is visible as a square covering the CNT in the region of the gate electrodes. Electrical transport and noise measurements were performed using SG1 and TG1. The scale bar corresponds to 500 nm.

Additionally, a CNT-FET structure with only a $SiO_2$ dielectric layer was investigated in order to compare the PLD-deposited $Al_2O_3$ and thermally grown $SiO_2$ as gate dielectrics. Here, the highly doped Si substrate was used as the back gate. The data obtained allowed us to separately determine the traps related to $SiO_2$ and $Al_2O_3$ dielectric layers.

To tune the characteristics of the FETs, they were exposed to small doses of gamma radiation. This treatment uses a standard isotope $^{60}$Co source, which emits characteristic gamma rays with a flux of 1 Gy/s and an energy of 1.2 MeV. The devices were treated with a dose of $10^4$ Gy, which is low in comparison to previous experiments with CNTs.[22]

Low-frequency noise spectra are recorded before and after gamma irradiation with a frequency ranging from 1 Hz to 100 kHz for different gate voltages in vacuum at a pressure of ~$10^{-4}$ mbar. The investigated temperature range was between 70 K and 290 K.

For the noise experiments, the samples were connected in series with a load resistance $R_{load}$ during data acquisition. With this setup, we calculated the current through the sample and observed the voltage fluctuations at the drain and source contacts of the FETs. The drain-source voltage $V_{DS}$ and the total voltage $V_M$ were measured to obtain the total current $I$ through the CNT channel. A small $V_{DS} = 30$ mV was chosen for all measurements. The value of $I$ was calculated by the following relation: $I = (V_M - V_{DS})/R_{load}$.

The noise signal from the sample was amplified by a home-made preamplifier with a gain of 177 and additionally by a commercial amplifier (ITHACO 1201) with a variable gain. The input impedance of the preamplifier in our experimental setup can be estimated as 1 MΩ in the AC regime (above 0.3 Hz). The input-related intrinsic thermal noise of the preamplifier and the ITHACO amplifier were measured as $2 \times 10^{-18}$ $V^2Hz^{-1}$ and $2 \times 10^{-17}$ $V^2Hz^{-1}$, respectively. The noise spectra, i.e. the voltage noise power spectral density $S_V$ as a function of frequency, were obtained using a dynamic signal analyzer (HP35670A).

The load resistance of 5 kΩ and 10 kΩ was used for different measurement series, which enabled us to operate in the short-circuit regime due to the high differential resistance, $R_{diff} = \partial V_{DS}/\partial I$, of the FET channel at each operating point. Taking into account the parallel connection of the resistances in the alternating current regime, the current noise power spectral density, $S_I$, is approximated by: $S_I \cong S_V/R^2_{load}$.

We used the normalized current noise power spectral density, $fS_I/I^2$, where $f$ is the frequency, in order to compare the noise components of FETs. This is a commonly used parameter to describe the noise performance of electrical devices since it is dimensionless and for the $1/f$-noise component it is independent of the frequency.

## III. RESULTS AND DISCUSSION

The transport and noise characteristics of the FETs were measured before and after gamma irradiation treatment. As was mentioned above, high doses of more than $5 \times 10^4$ Gy were previously used to tune the characteristics of the CNT-based devices and even modify the properties of the CNTs themselves. In this work we study the influence of a small dose of gamma radiation on the electrical characteristics of CNT-FET devices with different gate geometries and dielectric layers limiting the dose to $1 \times 10^4$ Gy. We investigate the transport and noise properties of FETs before and after gamma irradiation with side gate (SG) and top gate (TG) configurations (Fig. 1). Here we describe the transport and noise properties of the CNT-FET concentrating on the side gate labeled SG1 and the top gate labeled TG1, while similar results were obtained on other gates. We measured the drain current, $I$, dependence on gate voltage, $V_G$, in a wide temperature range from 70 K to 280 K before and after gamma radiation treatment (Fig. 2). The data show that in the on-state of the FET, at negative gate voltages, the current through the FET channel increases with temperature. The transfer characteristics of the FET show the improved control of the transistor and an increase of the total current in the on-state of the transistor after gamma irradiation. The maximum transconductance before irradiation was measured as $1.1 \times 10^{-7}$ S and $1.8 \times 10^{-7}$ S, for top gate and side gate, respectively. After irradiation, the transconductance increased to $1.5 \times 10^{-7}$ S and $2.1 \times 10^{-7}$ S, respectively. This can be explained by an improved interface between the electrodes (drain and source) and the CNT due to a rearrangement of atoms at the Pd-CNT interface as a result of their radiation-stimulated diffusion.[46] Furthermore, the side gate has a stronger impact on the FET conductance compared to the top gate. This is reasonable, since the side gate affects a larger area of the CNT. In addition, different quantities of active traps can be registered in the noise spectra depending on operation of the FET with the top gate or with side gate configuration due to a different potential profile along the CNT channel as it will be shown below in section IIIb.

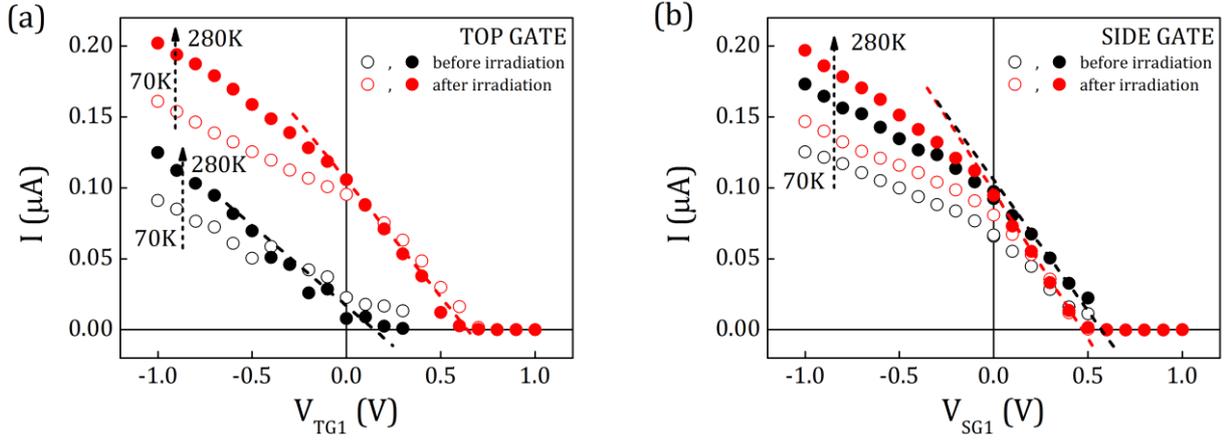

Fig. 2. Transfer characteristics of the CNT-FET before and after gamma irradiation measured in a wide temperature range from 70 K to 280 K (shown only for two temperatures) at different (a) top gate voltages and (b) side gate voltages. The radiation treatment results in an increase of the current in the on-state and an increase of the maximum value of transistor transconductance. The latter is reflected by the slopes of dashed lines shown for $T$ = 280 K (black line – before irradiation and red – after irradiation). $V_{DS}$ = 30 mV.

Noise spectra were measured at different points of the transconductance curves (Fig. 2) before and after gamma irradiation to analyze the transport properties of the CNT-FETs, including the influence of the contacts on the conductivity in the system. Using Pd as contact material results in nearly tunnel-transparent contacts to CNTs due to the large electron work function of this metal. In addition, this leads to near-ohmic behavior of the contacts for semiconducting nanotubes with diameter >1.6 nm.[13] Nevertheless, as will be shown below, our results demonstrate that the influence of the Schottky barriers at the Pd/CNT interfaces needs to be taken into account when operating the FET in the near-threshold regime.

The noise spectra of the CNT-FET before gamma irradiation will now be discussed in more detail. Typical noise spectra measured at different temperatures using the top gate are shown in Fig. 3 (similar spectra were obtained for the side gate SG1). The measurements were performed for the operating points around the maximum transconductance at voltages between $V_G$ = -0.5 V and 0 V. The noise spectra reveal several GR noise components in addition to flicker noise. As an example of our fitting procedure, we show the resulting fit for the noise spectrum measured at $T$ = 90 K (red dashed line). It consists of one flicker noise component (blue line), $S_V \sim f^{-1}$, and two GR noise components (green dashed lines), $S_V^{(1)} = S_{V0}^{(1)}/(1 + (f/f_1)^2)$ and $S_V^{(2)} = S_{V0}^{(2)}/(1 + (f/f_2)^2)$, where $S_{V0}^{(1)}$ and $S_{V0}^{(2)}$ – are the plateaus of the GR noise components, and $f_1$ and $f_2$ – are their characteristic frequencies. The resulting fit of $S_V$ is converted into the more convenient plot of the normalized current power spectral density. The data demonstrate that characteristic frequencies clearly depend on temperature. Later we also use the frequencies to analyze the resulting characteristic times $\tau = 1/(2\pi f_0)$, where $f_0$ is the characteristic frequency of each GR noise component, to analyze the activation energies of the charge traps involved in the generation of the GR noise component.

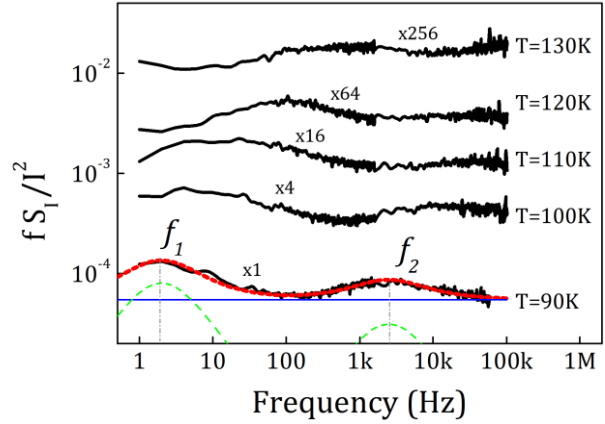

Fig. 3. Normalized current noise power spectral density measured at different temperatures for top gate before gamma irradiation at the operating points with maximum transconductance. The measured data are shifted upwards for clarity (factors are indicated near each curve). The red dashed curve is the fit of $f\,S_I/I^2$ by combining a flicker noise component (blue line) and two GR noise components (green dashed lines) with characteristic frequencies $f_1$ and $f_2$.

The GR noise components were clearly present before the gamma irradiation. At T=120 K, we find three GR noise components GR1 to GR3 (red lines in Fig. 4a). Their characteristic time constants $\tau$ are almost independent of the top gate voltage (Fig. 4b), while their amplitudes, $f_0 S_{I0}/2I^2$, in the noise spectra (inset in Fig. 4b) change significantly when the gate voltage is varied. Similar results have been obtained for the side gate geometry (not shown here). The dependences on gate voltage can be explained by changing the total number of charge carriers (and a small shift of the Fermi level) in the channel.[41,47] This fact has to be taken into account for the analysis of the GR noise components, in particular for deriving the activation energy of the charge traps causing GR noise, as we will show in the last part of the paper.

After gamma irradiation, the GR noise component vanished completely even for low numbers of carriers (near the closed state of the transistor), where in the case before irradiation the GR processes are clearly visible due to the strong influence of active traps and the low carrier concentration in the channel. Normalized current noise spectral densities measured at a temperature of 120 K and at different gate voltages, i.e. at different carrier concentrations, are shown in Fig. 4a after gamma irradiation (black curves) in direct comparison to the spectra before irradiation (red curves). It is apparent that almost all noise spectra after irradiation can be described by the flicker noise component alone and do not exhibit any features that can be ascribed to GR processes. This shows that the intensity of trapping/detrapping processes decreases after irradiation indicating a reduced number of active traps in the dielectric material close to the CNT transport channel. In addition, the direct comparison of the noise spectra before and after gamma irradiation reveals a decrease of the flicker noise component (down-shift of the normalized flicker noise-component in Fig. 4a).

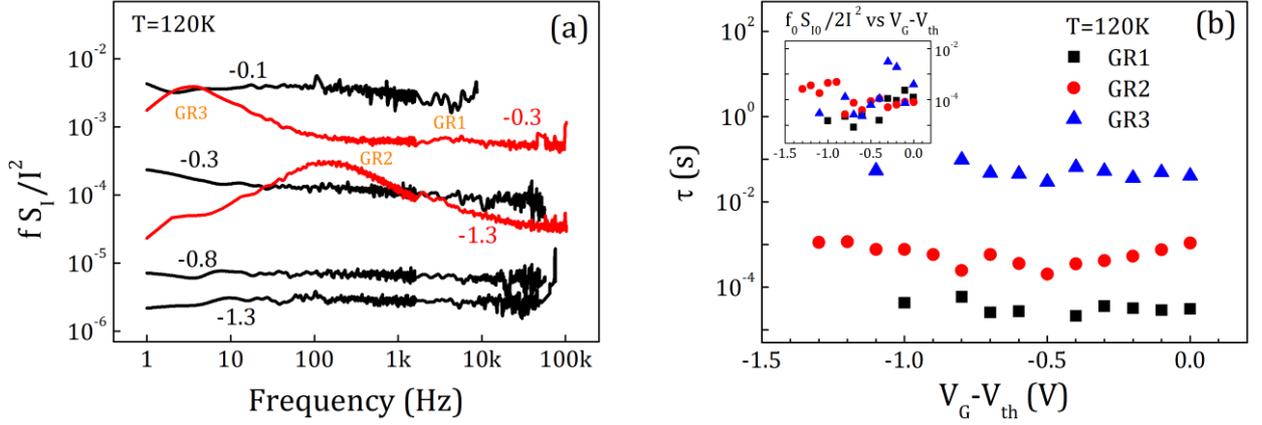

Fig. 4. (a). Normalized current noise power spectral density of the CNT-FET measured at different top gate overdrive voltages, $V_G - V_{th}$, (indicated near each of curves in V), and at $T$=120 K before (red) and after (black) gamma irradiation. (b) Characteristic times of GR noise components extracted from measured spectra at different gate voltages and at $T$=120 K. Inset: peak values of normalized current noise power spectral density of GR noise components (from panel a), as a function of ($V_G - V_{th}$).

Similar to the GR noise components, the amplitude of the normalized flicker noise component changes with gate voltage. The normalized current noise power spectra increase when the FET is near the off-state for the spectra taken both before and after gamma irradiation. This behavior can be described by the Hooge relation[31]:

$$\frac{fS_I}{I^2} = \frac{\alpha_H}{N}, \qquad (1)$$

where $\alpha_H$ is the Hooge parameter and $N$ is the number of carriers. Although we used Pd as contact material, reported in the literature as transparent barriers at the Pd/CNT interface[13], the increase of the flicker noise component with increasing $V_G$ reveals a strong impact of Schottky barriers forming at the interface in the near off-state regime of CNT-FET operation. Their influence on the transport properties of the FET will be discussed in the following section.

### IIIa. *Analysis of the impact of Schottky barriers on transport and noise properties*

As described above, Pd metallization was used to form almost transparent Schottky barriers with the CNT. However, the applied gate voltage changes the energy diagram increasing or decreasing the barriers for charge carriers flowing through the contact regions. In the present case, the Schottky barriers are transparent at $V_G$ =0 V due to the choice of the contact material, the small band gap of the CNT, and the operation of the CNT in the hole-regime (see Fig. 2).

In Fig. 5, we compare the gate voltage dependence of the flicker noise component at different temperatures before (Fig. 5a) and after (Fig. 5b) gamma irradiation. In the on-state of the FET, the flicker noise component is almost constant with the gate voltage $V_G$. In the off-state the flicker noise component increases exponentially for $V_G > \sim 0.2$ V. This reveals the strong influence of the Schottky barriers on the flicker noise component.

Due to the gamma irradiation, the flicker noise component is reduced by a factor of 1.7 from $3.7 \times 10^{-6}$ $f S_I/I^2$ to $2.1 \times 10^{-6}$ in the on-state of the FET (compare solid black lines in Figs. 5a and 5b). This can be attributed to an increased number of charge carriers (see Eq. (1)) and is in good agreement with a reduced number of charge traps as well as with the increased total current after radiation treatment.

Furthermore, the increase of the flicker noise component in the near off-state is much stronger after irradiation than before treatment. This indicates that the interfaces between the Pd metal contacts and the CNT are improved and hence the impact of the Schottky barriers on the transport is reduced. Thus, small doses of gamma irradiation lead to an overall improvement in the operation and noise characteristics of the CNT-FETs.

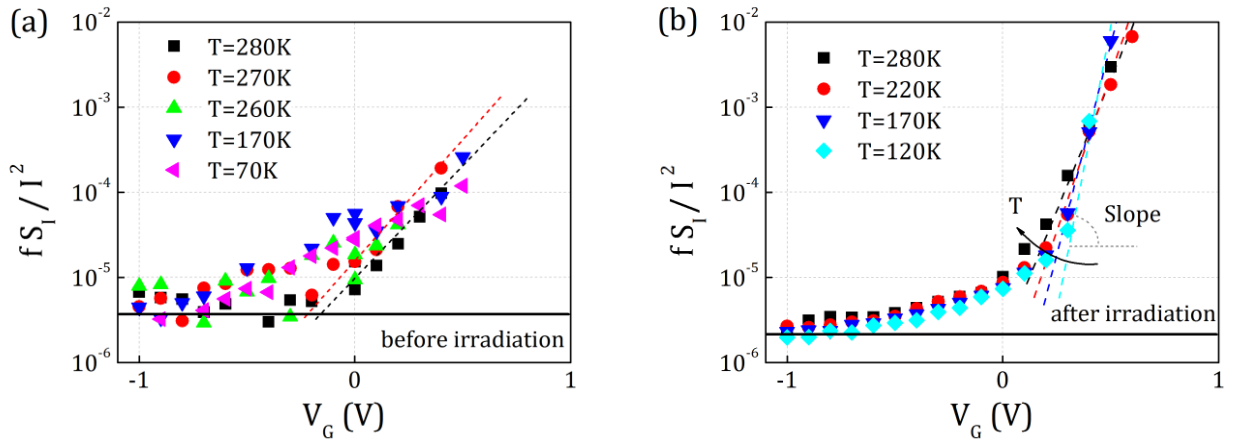

Fig. 5. Normalized current noise spectral density of the flicker noise component measured at different temperatures for the side gate geometry: (a) before and (b) after gamma irradiation. The slope of the fitted dashed lines depends on temperature. The horizontal black lines indicate the different values of flicker noise in the on-state of the CNT-FET.

Fig. 5b shows that the slope in the region at $V_G > 0.2$ V is steeper for lower temperatures. This behavior can be used to estimate the influence of the gate on the Schottky barrier, as discussed in the following. The width of the Schottky barriers increases with positive gate voltage. This results in a decreased hole current. Changing the gate voltage modulates the space-charge region at both Pd-CNT interfaces of the FET device. The current, $I_b$, through the Schottky barriers displays exponential behavior as a function of the voltage $V_b$ applied to the barrier, $I_b \sim T^2 \exp\left(-\frac{eV_b}{kT}\right)$, where $T$ is the temperature, $e$ is the electron charge and $k$ is the Boltzmann constant. In CNT-FET devices, $V_b$ depends on the applied drain-source voltage, $V_{DS}$, and on the gate voltage, $V_G$. Applying a small $V_{DS}$ = 30 mV hardly changes the barriers at all. Keeping $V_{DS}$ constant, we obtain the current $I$ through the CNT channel:

$$I \sim T^2 \exp\left(-\frac{erV_G}{kT}\right), \quad (2)$$

where $r$ is the lever arm of the gate voltage acting on the barriers. Using Eq. (1) and assuming that the Hooge parameter $\alpha_H$ does not depend on the gate voltage we obtain:

$$\frac{fS_I T^2}{I^2} \sim \exp\left(\frac{erV_G}{kT}\right). \quad (3)$$

The coefficient $r$ can be found by extracting the slope of the dependence of $\ln(fS_I/I^2)$ on $V_G$ (Fig. 5) and plotting it versus the inverse temperature. The results are shown in Fig. 6 for the side gate as well as for the top gate configurations. We extract $r$ from the linear fit of the data points using:

$$\frac{\Delta \ln(\frac{fS_I}{I^2})}{\Delta V_G} \sim \frac{re}{k}\frac{1}{T}. \quad (4)$$

Thereby, we find $r_{TG}$ is equal to 0.07±0.01 and $r_{SG}$ = 0.27±0.04 for the top and the side gates, respectively.

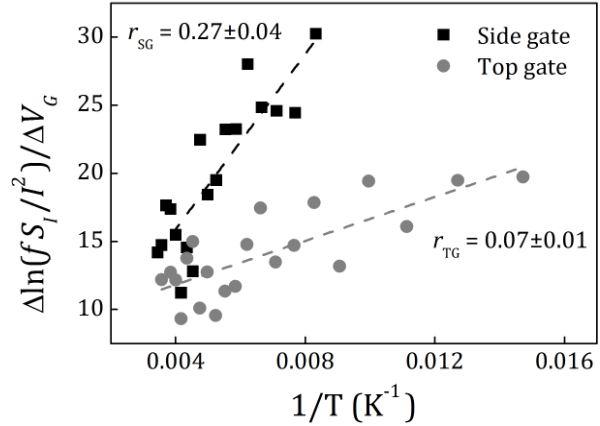

Fig. 6. Slope of the flicker noise component extracted from the $\ln(fS_I/I^2)$ dependence on $V_G$ plotted as a function of the inverse temperature for the top (TG) and the side (SG) gate after gamma irradiation. Data points are fitted linearly (dashed lines) and values for the coefficient $r$ resulting from these fits are given.

The values obtained for the coefficient $r$ show that the influence of the gate voltage on the Schottky contact region is stronger for the side gate configuration than for the top gate configuration. This observation can be explained by the different distances of the gates to the CNT. The top gate is approximately ten times closer to the channel in our geometry. Therefore, it controls only a small region of the nanotube directly underneath the gate while the Schottky barriers remain unchanged. This is in contrast to the side gate, which has a direct impact on the width of the Schottky barriers. It should be noted that it was possible to analyze the influence of the different gate configurations on the Schottky barriers for the CNT-FETs only due to the positive influence of gamma radiation on their functionality. In the case of samples measured before gamma radiation treatment, too high GR noise components in the spectra prevented obtaining data for such an evaluation of the FETs.

### IIIb. *Activation energies of active traps in Al$_2$O$_3$ and SiO$_2$ dielectric layers.*

As mentioned above, the temperature dependence of the GR noise components (Fig. 3) before gamma irradiation can be used to analyze the activation energies of the charge traps involved in the generation-recombination processes. Noise spectra similar to those presented in Fig. 3 were measured in a wide temperature range with steps of 2 K. The number of GR noise components varies with temperature, and it was possible to identify up to four GR noise components in each of the spectra. Noise spectra were measured at the gate voltages corresponding to maximum transconductance to ensure that the sample is in the same condition at different temperatures. All components are shifted to higher frequencies with increasing temperature until they vanish in thermal noise due to decreasing amplitudes. Additionally, new GR components were registered close to room temperature in the low-frequency range.

We identify several traps with different activation energies by calculating the characteristic time, $\tau$, of the GR process for each GR component from its characteristic frequency, $f$, according to $\tau = 1/(2\pi f)$ and plotting $\tau T^2$ as a function of 1000/T in a semi-logarithmic scale (Arrhenius plot). Then, the activation energy of the charge traps can be evaluated from the slope of a linear fit of the corresponding GR component, as shown in Fig. 7.

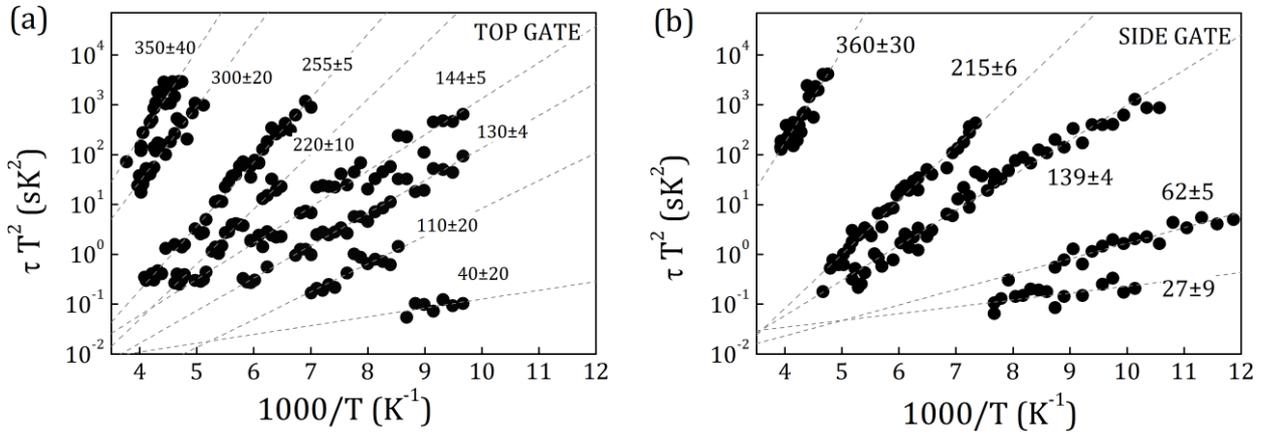

Fig. 7. Arrhenius plots obtained from measured GR noise components at different temperatures, *T*, before gamma irradiation of the FET at the gate voltage corresponding to the maximum transconductance using (a) the top gate and (b) the side gate. The value of the activation energies (in meV) of charge traps obtained from a linear fit (dashed lines) together with estimated errors is given next to each of the fits.

The activation energies of traps were found to be around 30 meV (27 meV for the side gate and 40 meV for the top gate), 62 meV, 130-144 meV, 220 meV, 255 meV, 300 meV and 350-360 meV. The value 110 meV can differ from the estimated value because the corresponding GR noise component has a small amplitude and almost overlaps with other GR components in the noise spectra. These activation energies are related to the traps near the interface between the CNT and the dielectric layer. In the CNT-FET, the nanotube is sandwiched between $Al_2O_3$ (top dielectric) and $SiO_2$ (bottom dielectric). Therefore, the obtained energies can be associated with both oxides in the CNT-FET sample. However, data obtained for the samples with a single $SiO_2$ dielectric layer allow us to separate the active traps belonging to each of the dielectric layers, as will be shown below and is summarized in Table 1.

Before that, we now turn to an analysis of the special features revealed in Fig. 7 and discuss the correlation between the number of traps obtained from the noise spectra and different gate geometries. Operating the FET with the top gate, more active traps can be registered in the noise spectra compared to the operation with the side gate. This observation can be explained by the following consideration. The localized electric field of the top gate may electronicaly separate the CNT into two parts (on each side of the gate electrode) with Fermi levels which are shifted by the drain-source voltage (30 mV) applied to the CNT source-drain contacts. In this case, the drain-source voltage drops mainly at the highly resistive central part, which is formed by applying a small positive gate voltage. Then, the energy difference between the Fermi level and the charge trap located in the dielectric layer near the CNT may vary for the two CNT parts on either side of the top gate (Fig. 8). Thus each charge trap level observed in the side gate operation splits into two levels in top gate operation due to a different potential profile along the CNT channel. Additionally, in this case the traps with small activation energies can hardly be recognized because of the relatively large number of GR noise components in the spectra measured in the low temperature range.

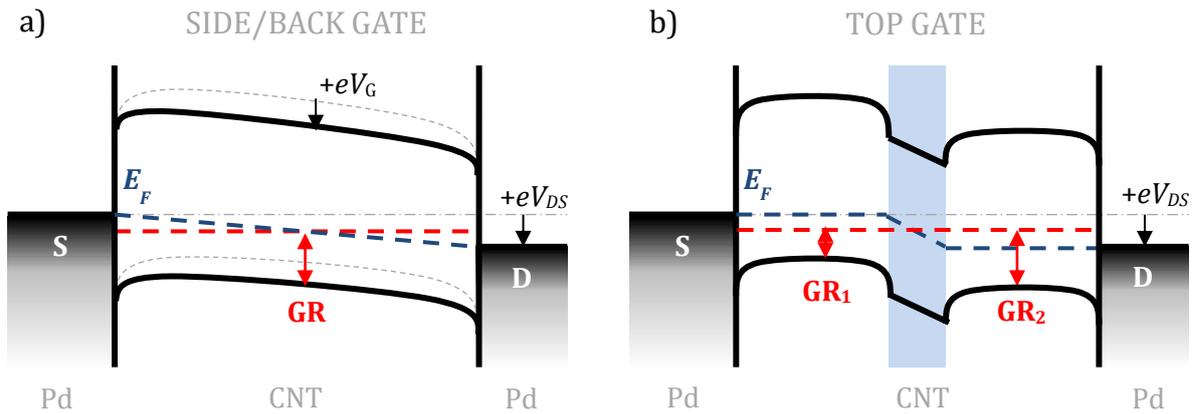

Fig. 8. Schematic energy band diagrams of a CNT-FET contacted by Pd electrodes and with (a) side gate and (b) top gate configuration. The potential profile along the CNT channel differs for side and top gate cases. A single trap level is assumed and sketched in red at a certain relative energy to the Fermi energy $E_F$. The side/back gate voltage in panel (a) lowers the CNT energy bands by $eV_G$ uniformly along the whole CNT channel. The activation energy GR is observed in the noise spectra. In panel (b), the top gate voltage separates the energy bands of the CNT into two parts. Two different activation energies $GR_1$ and $GR_2$ are observed for the same trap level.

In order to assign the observed charge traps to one of the dielectric layers, we measured the CNT-FETs fabricated using the same technology but with only one $SiO_2$ dielectric layer. As in the case of two dielectric layers, the GR noise components were clearly visible in the measured spectra. Their analysis revealed several types of traps with different activation energies (Fig. 9). The energies were found to be around 63 meV, 140 meV, 214 meV and greater than 280 meV.

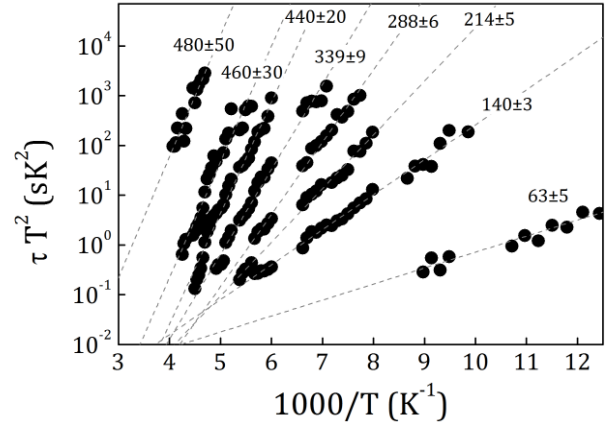

Fig. 9. Arrhenius plot on the basis of the noise measurement data obtained for the CNT-FET device with a single $SiO_2$ dielectric layer at a gate voltage corresponding to the maximum transconductance and at $V_{DS} = 30$ mV. The value of the activation energies (in meV) of charge traps obtained from a linear fit (dashed lines) together with estimated errors is given next to each of the fits.

The activation energies of traps obtained for the investigated samples are listed in Table 1. Comparing the two samples, the activation energies around 255 meV and ~30 meV can be attributed to traps in the $Al_2O_3$. Traps with activation energies of 65 meV (the trap 110 meV may be overrated because of a relatively large error due to its close location to the trap with an energy of 130 meV, which makes it difficult to extract the exact energy value in this case), 130-144 meV and 215-220 meV correspond to $SiO_2$. High activation energies with values above 400 meV, obtained in the case of CNT-FET with a single $SiO_2$ layer (see Fig. 9), can have the following origin. These active traps in the $SiO_2$ may be registered only when the shallower traps of the $Al_2O_3$ are not introduced into the system, i.e. in the CNT-FET structure without an $Al_2O_3$ dielectric layer.

Table 1. The activation energies, $E_A$, calculated from Fig. 7 for CNT-FETs with $SiO_2$ and $Al_2O_3$ dielectric layers and for CNT-FETs with a single $SiO_2$ layer (Fig. 9). It is also stated to which dielectric layer the traps can be ascribed to.

| Dielectric layers | | $E_A$ Related to |
|---|---|---|
| $SiO_2+Al_2O_3$ $E_A$ (meV) | $SiO_2$ $E_A$ (meV) | |
| - | 480, 460, 440 | $SiO_2$ |
| 360, 350, 300 | 339, 288 | $SiO_2$ |
| 255 | - | $Al_2O_3$ |
| 215, 220 | 214 | $SiO_2$ |
| 144, 139, 130 | 140 | $SiO_2$ |
| 110, 62 | 63 | $SiO_2$ |
| 40, 27 | - | $Al_2O_3$ |

The results obtained demonstrate the contribution of charge traps of both dielectric layers to the noise characteristics of CNT-FETs and to the device performance. This has to be taken into account for optimizing device performance, because such active traps determine the reliability and the stability of FETs.

### IIIc. *Intrinsic CNT transport properties determined using noise spectroscopy.*

The normalized current noise power spectral density is almost independent of the gate voltage at relatively high negative voltages when the FET is in its on-state (see Fig. 5a,b). In this operating regime, the Schottky barriers are thinner and the main source of flicker noise is the transport channel of the transistor. Thus, we can use this regime to characterize the intrinsic transport properties of the CNT itself. For small drain-source voltages the density of the current through the channel can be estimated as follows:

$$\frac{I}{S} = e \frac{N}{L_{DS}S} \mu \frac{V_{DS}}{L_{DS}}, \quad (5)$$

where $I$ is the drain current, $L_{DS}$ is the length of the CNT, $\mu$ is the carrier mobility, and $S$ is the radial section area of the CNT. We can obtain the number of carriers in the CNT by:

$$N = \frac{IL_{DS}^2}{V_{DS}e\mu}. \quad (6)$$

By substituting the number of carriers from Eq. (6) in the Hooge relation (1), we obtain:

$$f\frac{S_I}{I^2} = \frac{\alpha_H V_{DS} e \mu}{IL_{DS}^2}. \quad (7)$$

Taking into account that the Hooge parameter reflects the difference in scattering mechanisms, the parameter can be estimated as:[30]

$$\alpha_H = 2 \times 10^{-3} \left(\frac{\mu}{\mu_{latt}}\right)^2, \quad (8)$$

where $\mu_{latt}$ is the value of the mobility if lattice scattering is the only scattering mechanism present in the system. By substituting Eq. (8) into Eq. (7) we obtain:

$$f\frac{S_I}{I^2} = 2 \times 10^{-3} \frac{V_{DS}e}{\mu_{latt}^2 IL_{DS}^2} \mu^3 \quad (9)$$

Taking into account that the intrinsic mobility of CNTs, $\mu_{latt}$, was estimated to be $10^5$ cm$^2$V$^{-1}$s$^{-1}$ at room temperature[48] and using our measured values of the power spectral density $S_I$, we obtain the value of the mobility $\mu$ to be around 26000 cm$^2$V$^{-1}$s$^{-1}$ and 22000 cm$^2$V$^{-1}$s$^{-1}$ at room temperature before and after irradiation, respectively. Such high mobility values indicate, first of all, the high quality of the CNT used in our investigations. Second, the very small decrease of the mobility after gamma irradiation reflects the radiation hardness of the structure and proves that we have found a radiation dose which can improve the FET properties without damaging the transport channel.

## IV. CONCLUSION

To summarize, individual carbon nanotube FETs with top, side, and back gate configurations and different dielectric layers were investigated with respect to their noise and transport properties before and after gamma radiation treatment. All devices showed a strong contribution of GR noise in the spectra before treatment. Several active charge traps with different activation energies were identified on the basis of the temperature dependence of the corresponding GR noise components. Moreover, it was revealed that in the noise spectra of the FET operated with the top gate, more active traps can be observed than in the case of the FET operated with side gates due to a different potential profile along the CNT channel. Comprehensive analysis of the data obtained for CNT-FETs with two dielectric layers, SiO$_2$ and Al$_2$O$_3$, and a single SiO$_2$ layer allowed us to attribute activation energies of 30 meV and 255 meV to traps within the Al$_2$O$_3$ layer, used to cover the CNT with side gate and top gate configurations.

After gamma irradiation with a dose of $1\times10^4$ Gy, all GR noise components vanish in spite of the reduced level of flicker noise, which is a clear indication that the number of charge traps decreases considerably. This is consistent with a reduction of the flicker noise component in the on-state of the FET, and, thus, an increased number of charge carriers in the channel after irradiation (for a constant Hooge parameter $\alpha_H$). The mobility derived from the flicker noise component in the on-state is greater than $2\times10^4$ cm$^2$V$^{-1}$s$^{-1}$ at room temperature. It decreases only slightly after radiation treatment demonstrating high radiation hardness of the device. Analyzing the flicker noise component in the regime near the off-state of the FET, we find that the Schottky barriers at the metal-nanotube interface are reduced as a result of gamma irradiation. Using noise spectroscopy, we find that the transport as well as the noise properties in a CNT-FET working in the on-state regime are determined by the CNT itself rather than the contact interface. This allows us to estimate the carrier mobility for the CNT channel.

We conclude that gamma irradiation applied to CNT-FETs reduces the contact barriers and passivates the charge traps within the dielectric layer and therefore increases the number of charge carriers in the channel. Thus, radiation treatment can be used to improve the overall performance of carbon nanotube field effect transistors.

**Acknowledgements.**

We thank S. Trellenkamp for e-beam writing and J. Schubert for PLD deposition. We gratefully acknowledge funding by the DFG within Research Unit 912 and support by the JARA Seed Fund. J. Li would like to acknowledge support from the China Scholarship Council.